\documentclass[aps,prl,prabib,epsfig,amsmath,amssymb,twocolumn,showpacs]{revtex4}
\usepackage{amsmath}
\usepackage{amssymb}
\usepackage{epsfig}
\usepackage{graphics}

\newcommand \beq{\begin{eqnarray}}
\newcommand \eeq{\end{eqnarray}}
\def\simge{\mathrel{%
       \rlap{\raise 0.511ex \hbox{$>$}}{\lower 0.511ex \hbox{$\sim$}}}}
\def\simle{\mathrel{
       \rlap{\raise 0.511ex \hbox{$<$}}{\lower 0.511ex \hbox{$\sim$}}}}

\begin{document}

\title{Breathing mode of two-dimensional atomic Fermi gases in harmonic traps}

\author{Chao Gao}
\author{Zhenhua Yu}
\email{huazhenyu2000@gmail.com}
\affiliation{Institute for Advanced Study, Tsinghua University, Beijing 100084, China}


\begin{abstract}

For two-dimensional (2D) atomic Fermi gases in harmonic traps, the $SO(2,1)$ symmetry is broken by the interatomic interaction explicitly via the contact correlation operator. Consequently the frequency of the breathing mode $\omega_B$ of the 2D Fermi gas can be different from $2\omega_0$, with $\omega_0$ the trapping frequency of harmonic potentials.
At zero temperature, we use the sum rules of density correlation functions to yield upper bounds for $\omega_B$. We further calculate $\omega_B$ through the Euler equations in the hydrodynamic regime. The obtained value of $\omega_B$ satisfies the upper bounds and shows deviation from $2\omega_0$ which can be as large as about $8 \%$. 
\end{abstract}

\maketitle

Collective oscillation modes of atomic gases confined spatially in harmonic traps
convey crucial information of the nature of the systems. The measurements of the frequency of the breathing mode in three dimensions have confirmed the superfluid hydrodynamics of Bose-Einstein condensates \cite{jin} and the universality of unitary Fermi gases \cite{thomas}. Moreover, it is pointed out in Ref.~\cite{pitaevskii} that if the interatomic interaction satisfies scale invariance, the atomic gases possess a ``hidden" $SO(2,1)$ symmetry. Such a symmetry dictates that the frequency of the breathing mode of the gases confined in a spherical harmonic potential with trapping frequency $\omega_0$ must be $2\omega_0$. 

With the advent of experimental realization of two-dimensional atomic gases both of bosons \cite{dalibardnat, dalibard, cornell, chin} and fermions \cite{turlapov, kohlnat, kohl, zwierlein, hu}, Refs.~\cite{olshanii} and \cite{hofmann} put into the context of quantum anomaly the point mentioned in Ref.~\cite{pitaevskii} that if a contact pseudo-potential with a bare coupling constant $\bar g$ is used to model the short-ranged interatomic interaction, the two-dimensional atomic gases have the $SO(2,1)$ symmetry at the classical level since $\bar g$ is dimensionless. However, necessary renormalization of $\bar g$ introduces the scattering length $a_{2D}$ in two dimensions  as a new low energy observable which characterizes the interatomic interaction. The classical $SO(2,1)$ symmetry ceases to hold at the quantum level due to the fact that $a_{2D}$ carries dimension. Consequently, the frequency of the breathing mode $\omega_B$ of the two-dimensional gases can be different from two times the harmonic frequency $\omega_0$ of the spherical traps.  On the other hand, recent experiment of two-dimensional Fermi gases carried out at temperature $T\approx0.37 T_F$, with $T_F$ the Fermi temperature, did not observe substantial deviation of $\omega_B$ from $2\omega_0$ at all \cite{kohlbm}.

\begin{figure}
\includegraphics[width=8.5cm]{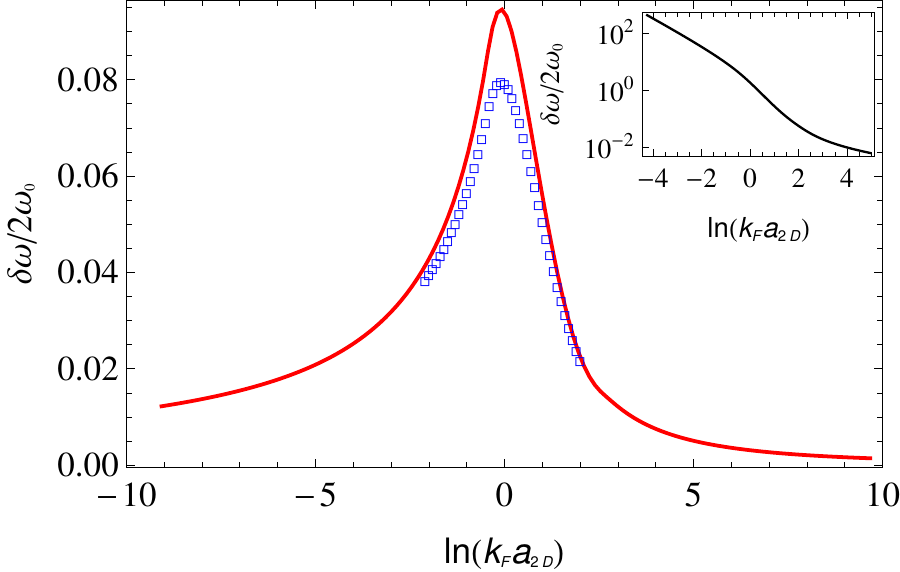}
\caption{(Color online) Relative deviations from $2\omega_0$ versus $\ln(k_F a_{2D})$ for the upper bounds $\omega_{1,-1}$ and $\omega_{3,1}$, and $\omega_B$ calculated by the hydrodynamic approach. The red curve is for $\delta\omega=\omega_{1,-1}-2\omega_0$ evaluated from Eq.~(\ref{up1}). The black curve in the inset is for $\delta\omega=\omega_{3,1}-2\omega_0$ from Eq.~(\ref{up2}). The blue squares are for $\delta\omega=\omega_B-2\omega_0$ evaluated from Eq.~(\ref{obhd}) for $-2<\ln(k_F a_{2D})<2$.} \label{f1}
\end{figure}

In this paper, we study the breathing mode of a two-dimensional (2D) Fermi gas of equal number of two species of fermionic atoms at zero temperature. Based on the short-rangedness of the interatomic interaction and the correlation structure at short distances in dilute Fermi gases, we give an alternative derivation to show that the interaction violates the $SO(2,1)$ symmetry explicitly via the contact correlation operator, compared to the quantum field approach in Ref.~\cite{hofmann}.
We derive the sum rules of density correlation functions and use them to yield upper bounds for the frequency of the breathing mode $\omega_B$. We further calculate $\omega_B$ by the Euler equations in the hydrodynamic regime and find that the obtained $\omega_B$ satisfies the upper bounds and can be about $8\%$ bigger than $2\omega_0$ in the unitary regime as shown in Fig.~(\ref{f1}). Since in the high temperature limit atoms are noninteracting and the system obeys the $SO(2,1)$ symmetry trivially, $\omega_B=2\omega_0$. Our results at zero temperature determine the typical order of magnitude how large $\omega_B$ can deviate from $2\omega_0$ at finite temperatures. Our results have the prospect to be verified by future experiment deep in the degenerate regime.

\emph{Basic formalism.---}
Experiment produces two-dimensional Fermi gases by confining $^6$Li atoms of two hyperfine states in a three-dimensional harmonic trap with a large trapping frequency $\omega_z$ in the $z$ direction \cite{turlapov, kohlnat, kohl, zwierlein}. Under the circumstances that all other relevant energy scales are much smaller than $\omega_z$, the motion of the atoms in the $z$ direction is frozen; low energy dynamics occurs only in the $xy$ plane. The system is described by the Hamiltonian
\begin{align}
\hat H=&\hat H_0+\hat H_{ho},\nonumber\\
\hat H_0=&\sum_{i,\sigma=\uparrow, \downarrow}\frac{\mathbf {\hat p}^2_{i,\sigma}}{2m}+\sum_{i, j}U(|\mathbf {\hat r}_{i,\uparrow}-\mathbf {\hat r}_{j,\downarrow}|),\nonumber\\
{\hat H}_{ho}=&\frac12m\omega_0^2 {\sum_{i,\sigma=\uparrow, \downarrow}\mathbf {\hat r}^2_{i,\sigma}},
\end{align}
with the operators $\mathbf {\hat r}=\{\hat x,\hat y\}$, $\mathbf {\hat p}=\{\hat p_x,\hat p_y\}$, and $m$ the atom mass and $\omega_0$ the spherical trapping frequency in the $xy$ plane.
We denote the two species of the fermions by $\uparrow$ and $\downarrow$. The attractive interaction potential $U(r)$ has a short-range $r_0$ and gives rise to a shallow bound state with binding energy $E_b=-1/ma_{2D}^2$, with  the 2D scattering length $a_{2D}\gg r_0$. (We take $\hbar=1$ throughout.) Since we are only interested in the long wavelength physics, the intra-species interactions have been neglected due to the Pauli exclusion principle.

Experimentally the breathing mode can be excited by modulating the trapping frequency $\omega_0$ \cite{kohlbm}. At zero temperature, the breathing mode should bring about a sharp peak in the spectrum function of the density correlation
\begin{align}
\chi''(\omega)=\pi\sum_{f} |\langle f| \delta{\hat O}|g\rangle|^2\delta(\omega-E_f+E_g),\label{chi}
\end{align}
where $|g\rangle$ is the ground state, and $\hat O=\sum_{i,\sigma=\uparrow, \downarrow}\mathbf {\hat r}^2_{i,\sigma}$ and $\delta\hat O=\hat O-\langle\hat O\rangle$. We use $\langle\dots\rangle$ to denote the expectation value over $|g\rangle$.

The structure of $\chi''(\omega)$ can be analyzed by calculating the equation of motion of the operator $\hat O$:
\begin{align}
i\frac\partial{\partial t}{\hat O}(t)=[\hat O,\hat H]=\frac{2i}m{\hat D}. \label{emo}
\end{align}
The scale dilation generator is $\hat D=\sum_{i,\sigma}(\mathbf {\hat p}_{i,\sigma}\cdot\mathbf {\hat r}_{i,\sigma}+\mathbf {\hat r}_{i,\sigma}\cdot\mathbf {\hat p}_{i,\sigma})/2$ since $e^{i\gamma{\hat D}} \mathbf {\hat r}_{i,\sigma} e^{-i\gamma{\hat D}}=e^{\gamma} \mathbf {\hat r}_{i,\sigma}$. We continue to calculate the equation of motion for $\hat D$: \begin{align}
i\frac\partial{\partial t}{\hat D}(t)=&[\hat D, \hat H]\nonumber\\
=&2i\hat H-4i\hat H_{ho}-i\int d^2\mathbf r\left[2U(r)+\frac{\partial U(r)}{\partial r}r\right]\hat\rho(\mathbf r).
\label{emd}
\end{align}
Here the two-particle correlation function is $\hat\rho(\mathbf r)=\sum_{i,j}\delta(\mathbf r-\mathbf {\hat r}_{i,\uparrow}-\mathbf {\hat r}_{j,\downarrow})$.
If the interatomic interaction is scale invariant, i.e., $U(e^{\gamma}r)=e^{-2\gamma}U(r)$,
the last term in Eq.~(\ref{emd}) is zero; given $[\hat D,\hat O]=-2i\hat O$,
the operators $\hat O$, $\hat D$ and $\hat H$ form
a closed algebra, which corresponds to a $SO(2,1)$ symmetry \cite{pitaevskii}. In this case, the operator $\hat O$ satisfies
\begin{align}
\frac{\partial^2}{\partial t^2}\hat O=\frac4m\hat H-4\omega_0^2\hat O,
\end{align}
from which one can read off that nonzero matrix elements $\langle f|\hat O(t)|g\rangle$ oscillate with a frequency $2\omega_0$ and conclude that $\chi''(\omega)$ is a delta function centering at the frequency of the breathing mode $\omega_B=2\omega_0$.

However, the real interatomic interaction is not scale invariant. To evaluate the last term in Eq.~(\ref{emd}) which breaks the $SO(2,1)$ symmetry, we
note that due to the diluteness of atomic Fermi gases the two-particle correlation function has the asymptotic form \cite{castin}
\begin{align}
\hat \rho(\mathbf r)=\hat {C}\phi^2(r)
\end{align}
for $r\ll d$ the mean interparticle spacing. The wave function $\phi(r)$ satisfies the two-body Schr\"odingier equation in the relative coordinates
\begin{align}
\left[-\frac1r\frac{d}{dr}r\frac{d}{dr}+mU(r)\right]\phi(r)=0\label{schrodinger}
\end{align}
and is normalized such that $\phi(r)=\ln(r/a_{2D})$ for $r\gtrsim r_0$ the range of $U(r)$.
The contact correlation operator $\hat {C}$, which quantifies the correlation strength between fermions at short distances, obeys the adiabatic relation \cite{castin}
\begin{align}
\langle{\hat C}\rangle=\frac m{2\pi}\frac{\partial \langle \hat H\rangle}{\partial \ln a_{2D}},\label{adia}
\end{align}
at zero temperature.
Similar to the manipulations employed in Ref.~\cite{zhang}, we integrate the second term of the integrand in Eq.~(\ref{emd}) by parts, and by Eq.~(\ref{schrodinger}) find
\begin{align}
\int d^2\mathbf r \left[2U(r)+\frac{\partial U(r)}{\partial r}r\right]\phi^2(r)=-2\pi/m.
\end{align}
Thus, we have
\begin{align}
[\hat D,\hat H]=2i\hat H-4i\hat H_{ho}+2i\pi\hat C/m,\label{com}
\end{align}
which agrees with the result derived by a quantum field approach in Ref.~\cite{hofmann}.

\emph{Sum rules.---}
With the contact correlation operator $\hat C$ breaking the $SO(2,1)$ symmetry explicitly, the breathing mode frequency $\omega_B$ can be different from $2\omega_0$.
To constrain the value of $\omega_B$, we use the sum rules, $s_{\ell}=\int_{-\infty}^{+\infty}d\omega\chi''(\omega)\omega^{\ell}/\pi$, to define the frequencies $\omega_{\ell,\ell-2}=\sqrt{s_{\ell}/s_{\ell-2}}$; at zero temperature $\omega_{\ell,\ell-2}$ are upper bounds for $\omega_B$ \cite{stringaribook}. Specifically we consider
\begin{align}
s_3=&\frac2{m^2}\langle [[\hat D,\hat H],\hat D]\rangle,\\
s_1=&\frac12\langle [[\delta\hat O,\hat H],\delta\hat O]\rangle=\frac2{m}\langle \hat O\rangle,\\
s_{-1}=&-\frac 1m\frac{\partial}{\partial\omega_0^2}\langle \hat O\rangle.
\end{align}
The third sum rule $s_3$ can be evaluated as
\begin{align}
&[[\hat D,\hat H],\hat D]\nonumber\\
&=4\hat H-{\hat C}\int d^2\mathbf r\left[4U(r)-r\frac{\partial}{\partial r}r\frac{\partial U(r)}{\partial r}\right]\phi^2(r).\label{dd}
\end{align}
To proceed further, we assume a square well model potential $U(r)=-V_0\theta(r_0-r)$ with $V_0>0$ and obtain
\begin{align}
&\int d^2\mathbf r\left[4U(r)-r\frac{\partial}{\partial r}r\frac{\partial U(r)}{\partial r}\right]\phi^2(r)\nonumber\\
&=-\frac{4\pi}m+4\pi V_0r_0^2\ln(r_0/a_{2D}).
\end{align}
On the other hand, since the parameters $V_0$ and $r_0$ are required to reproduce the low energy physical quantities, e.g., the scattering length $a_{2D}$, in the limit $r_0/a_{2D}\to0$, we have
\begin{align}
\ln(r_0/a_{2D})=-\frac2{m V_0 r_0^2}.\label{renorm}
\end{align}
From Eqs.~(\ref{dd})-(\ref{renorm}) we obtain
\begin{align}
s_3=\frac8{m^2}\left[\langle\hat H\rangle+\frac{3\pi}m
\langle\hat C\rangle\right],\label{s3}
\end{align}
which also implies $[\hat D,\hat {C}]=4i\hat{C}$
where the factor $4$ is the dimension of ${\hat C}/\Omega$ with $\Omega$ the system volume in two dimensions. 

By the virial theorem
\begin{align}
\langle \hat H\rangle=m\omega_0^2\langle \hat O\rangle-\frac\pi m \langle {\hat C} \rangle,
\end{align}
which can be derived from
\begin{align}
\left(\omega_0\frac{\partial}{\partial\omega_0}-\frac12\frac{\partial}{\partial\ln a_{2D}}-1\right)\langle \hat H\rangle=0,
\end{align}
we cast $\omega_{3,1}$ into the form
\begin{align}
\omega_{3,1}=2\omega_0\left[1+\frac{2\pi\langle\hat{C}\rangle}{m^2\omega_0^2\langle\hat O\rangle}\right]^{1/2}.\label{up1}
\end{align}
On the other hand, using dimensional analysis, we find within the local density approximation the upper bound
\begin{align}
\omega_{1,-1}=2\omega_0/\sqrt{1-\Delta} \label{up2}
\end{align}
with
\begin{align}
\Delta=\frac{\pi(2+\partial/\partial\ln a_{2D})\langle\hat C\rangle}{2m^2\omega_0^2\langle \hat O\rangle}.\label{delta}
\end{align}

We evaluate $\omega_{3,1}$ and $\omega_{1,-1}$ for the 2D Fermi gas trapped in the harmonic potential with the local density approximation and using the interpolation of the equation of state obtained by the Monte Carlo simulation \cite{giorgini} as used in Ref.~\cite{hu}. Figure (\ref{f1}) shows the relative deviations of the upper bounds $\omega_{3,1}$ and $\omega_{1,-1}$ from $2\omega_0$ verse $\ln( k_F a_{2D})$, where $k_F^2=2\pi n_0$ and $n_0$ is the total fermion density at the trap center.
In the BEC limit $a_{2D}\to0^+$, fermions pair to form tightly bound bosonic molecules and these molecules are weakly interacting with each other. The ground state energy is $\langle \hat H\rangle\approx -N/2m a_{2D}^2$ with $N$ the total number of fermions. From Eq.~(\ref{adia}), $\langle\hat C\rangle\approx N/2\pi a_{2D}^2$, while $\langle\hat O\rangle\sim N/\sqrt{m\omega_0}$.
The upper bound $\omega_{3,1}$ diverges as $a_{2D}\to 0^+$. The physical reason of this divergence is that when the trapping frequency $\omega_0$ is modulated, the operator $\hat O$ acting on $|g\rangle$ can disassociate the molecules. The integral of $\omega^3\chi''(\omega)$ is dominated by the part at frequencies $\omega\gtrsim1/ma_{2D}^2$ (cf.~Eq.~(\ref{chi})). In the BCS limit $a_{2D}\to+\infty$, the interaction energy of the ground state is $\sim-1/\ln (a_{2D})$, and thus $\hat C\sim1/\ln^2 (a_{2D})$, while $\langle\hat O\rangle$ is basically the value for noninteracting fermions which is finite; $\omega_{3,1}$ approaches $2\omega_0$ from above.

The other upper bound $\omega_{1,-1}$ does not suffer divergence in the BEC limit, which mathematically is due to lower powers of $\omega$ in $s_{1}$ than in $s_{3}$. On the ground of dimensional analysis, $\langle \hat O\rangle=f(m\omega_0a^2_{2D},N)/\omega_0$. We express $\omega_{1,-1}$ in terms of the function $f$ as
\begin{align}
\omega^2_{1,-1}=\left.\frac{4\omega_0^2 f(\xi,N)}{f-\xi \partial f(\xi,N)/\partial\xi}\right|_{\xi=m\omega_0a_{2D}^2}.
\end{align}
Since the cloud size increases as $a_{2D}$ increases \cite{hu}, $\partial f(\xi,N)/\partial\xi>0$; $\omega_{1,-1}$ is always bigger than $2\omega_0$, which implies $\Delta>0$. From the equation of state of homogeneous gases obtained in Ref.~\cite{giorgini}, one can deduce that $\xi\partial f(\xi,N)/\partial\xi$ becomes zero in the BEC and BCS limits and reaches a maximum at $\xi\sim1$. Correspondingly, as shown in Fig.~(\ref{f1}), the bound $\omega_{1,-1}$ approaches $2\omega_0$ in the BEC and BCS limits and shows a maximum in the unitary regime $k_F a_{2D}\sim1$. We find $\omega_{1,-1}<\omega_{3,1}$ for any $\ln(k_F a_{2D})$; the upper bound $\omega_{1,-1}$ is more strict.

\emph{Hydrodynamic equations.---}
The above sum rule results
suggest that the frequency of the breathing mode $\omega_B$ can differ from $2\omega_0$ by a significant amount in the unitary regime.  The experiment \cite{kohlbm} showed that the degenerate Fermi gas is in the hydrodynamic regime around the unitary limit $ k_F a_{2D}=1$. 
The small decay rate of the breathing mode measured there, which is primarily due to the anharmonicity of the trapping potential \cite{grimm}, indicates that dissipation is negligible. We use the Euler equations
\begin{align}
&\frac{\partial n}{\partial t}+\triangledown\cdot(n\mathbf v)=0,\label{eu1}\\
&m\left(\frac{\partial\mathbf v}{\partial t}+\mathbf v\cdot\triangledown\mathbf v\right)=-\triangledown\mu-\triangledown V_{ho},\label{eu2}
\end{align}
to calculate $\omega_B$ in the hydrodynamic regime at zero temperature. Here $n$ and $\mathbf v$ are the density and velocity fields of the fermions, and $\mu(\mathbf r)$ is the local chemical potential of the fermions. The harmonic potential is $V_{ho}(\mathbf r)=m\omega_0^2\mathbf r^2/2$.

Previous applications of the Euler equations to atomic gases in three dimensions usually linearize the density $n=n_{eq}+\delta n$ with $n_{eq}$ the equilibrium density distribution, and treat both $\delta n$ and $\mathbf v$  as small quantities. If one assumes $\delta n(\mathbf r,t)=e^{-i\omega t}\delta n(\mathbf r)$, the linearized Euler equations give \cite{pethick}
\begin{align}
-\omega^2\delta n=\triangledown\cdot\left[n_{eq}\triangledown\left(\frac{\partial\mu_{eq}}{\partial n_{eq}}\delta n\right)\right].\label{le}
\end{align}
Here $\mu_{eq}$ is the equilibrium local chemical potential.
It is tempting to solve Eq.~(\ref{le}) for the breathing mode of two-dimensional Fermi gases as an eigen-equation for $\delta n$ in the domain $0<r<R_{TF}$ with $R_{TF}$ the Thomas-Fermi radius of the cloud at equilibrium where $n_{eq}(R_{TF})=0$. However, due to the attractive interatomic interaction,
the part of the system close to the cloud edge is in the BEC regime. The equation of state for this part is $\mu-1/2ma_{2D}^2\sim-n/\ln(n a_{2D}^2)$ \cite{olshanii}. Given that $\delta n$ for the breathing mode should be spherically symmetric, the asymptotic form of Eq.~(\ref{le}) in this spatial region is
\begin{align}
\left(w\frac{d^2}{dw^2}+\frac{d}{dw}+\frac1{w\ln^2 w}\right)\delta n=0
\end{align}
with $w=1-r^2/R_{TF}^2$. We discover $\delta n\sim (-\ln w)^{1/2}$ which diverges as $w\to0^+$ \cite{bender}; no physical solution exists. Mathematically speaking, the logarithm in the equation of state of two-dimensional BECs makes $r=R_{TF}$ an irregular singular point of Eq.~(\ref{le}). We attribute the absence of physical solutions to Eq.~(\ref{le}) to the approximation of linearizing the Euler equations which must breaks down since the linearized solution $\delta n$ would become infinitely big close to the cloud edge.

Physically the frequency of the breathing mode $\omega_B$ is expected to be determined by the bulk of the cloud instead of the details close to the edge. To emphasize the property of the bulk,
we adopt a variational approach and work with the Lagrangian
\begin{align}
L=\int d^2\mathbf r\left[\frac12m n\mathbf v^2-\mathcal E-nV_{ho}+\phi\left(\frac{\partial n}{\partial t}+\triangledown\cdot(n\mathbf v)\right)\right], \label{lag}
\end{align}
from which Eqs.~(\ref{eu1}) and (\ref{eu2}) can be derived \cite{ed}. The internal energy density $\mathcal E$ is related to the local chemical potential $\mu$ via $\partial\mathcal E/\partial n=\mu$. The Lagrangian multiplier $\phi$ is used to ensure the continuity equation (\ref{eu1}). We assume a variational ansatz $n(\mathbf r,t)=n_{eq}(\mathbf r/\lambda(t))/\lambda(t)^2$ for the breathing mode which is exact if the interaction is scale invariant. Correspondingly Eq.~(\ref{eu1}) determines $\mathbf v(\mathbf r,t)=(d\ln\lambda(t)/dt)\mathbf r$. Substituting the forms of $n(\mathbf r,t)$ and $\mathbf v(\mathbf r,t)$ into Eq.~(\ref{lag}), for small oscillations $\lambda(t)=1+\delta(t)$ with $\delta(t)\ll1$, we obtain the Lagrangian equation
\begin{align}
d^2\delta(t)/dt^2+\omega_{B}^2\delta(t)=0,
\end{align}
with
\begin{align}
\omega_{B}^2=\frac4m\frac{\int d^2\mathbf r \;n_{eq}^2(\partial\mu_{eq}/\partial n_{eq})}{\int d^2\mathbf r \;n_{eq}\mathbf r^2}.\label{obhd}
\end{align}
Further manipulation yields 
\begin{align}
\omega_B=2\omega_0\sqrt{1+\Delta},
\end{align} 
from which one can tell $\omega_B<\omega_{1,-1}$ for $\Delta>0$ (cf.~Eq.~(\ref{up2})).
The physical meaning of Eq.~(\ref{obhd}) is manifest: its nominator is the average of the square of the sound velocity over the cloud and its denominator gives the square of the linear dimension of the cloud;
$\omega_B$ is basically the rate how fast a density variation can propagate through the cloud with the averaged sound velocity \cite{randeria}.
Note that the logarithmic singularity in the equation of state for two-dimensional BECs mentioned above is weak enough that the integral in the nominator of Eq.~(\ref{obhd}) converges.
If the interaction is scale invariant, $n_{eq}(\partial\mu_{eq}/\partial n_{eq})=\mu_{eq}$; given $\partial\mu_{eq}/\partial r=-\partial V_{ho}/\partial r$, one retrieves $\omega_{B}=2\omega_0$.

With the local density approximation as used for $\omega_{1,-1}$ and $\omega_{3,1}$, we evaluate $\omega_B$ from Eq.~(\ref{obhd}) for $-2<\ln(k_F a_{2D})<2$ which is presumably in the hydrodynamic regime \cite{kohlbm}. Figure (\ref{f1}) shows that the maximum deviation of $\omega_B$ from $2\omega_0$ is about $8\%$ at $k_F a_{2D}\sim1$. The estimate of $\omega_B$ in Ref.~\cite{hofmann}, where a linearized Euler equation is solved with assuming a polytropic equation of state, is  bigger than ours, and barely satisfies the upper bound $\omega_{1,-1}$.
Our calculation of $\omega_B$ at zero temperature sets up the typical scale how large the difference between $\omega_B$ and $2\omega_0$ can be. It is natural to expect that when temperature rises up, $\omega_B$ decreases from its value at zero temperature, and approaches $2\omega_0$ in the high temperature limit where the fermions are essentially noninteracting.
Future experimental advance into deep degenerate regimes has the prospect of suppressing the damping rate of the breathing mode, which is about $2\%$ of $2\omega_0$ at $T/T_F\approx0.37$ \cite{kohlbm}, and producing a definite detection of nonzero $\omega_B-2\omega_0$.

We acknowledge useful discussions with H.~ Zhai, Z.~Chen, G.~Baym, C.J.~Pethick, M.~K\"ohl, M.~Randeria, E.~Taylor, and M.~Oshanii. We thank M.~Oshanii for providing us with his notes on solving the Euler equations for two dimensional gases. This work is supported in part by Tsinghua University Initiative Scientific Research Program, and NSFC under Grant No. 11104157.

\end{document}